\documentclass[]{raa}
\usepackage[]{natbib,amsmath,amssymb}
\usepackage{graphicx,times,epsfig} 

\renewcommand{\d}{\mathrm{d}}

\newcommand{\ii}{\mathrm{i}}
\newcommand{\bea}{\begin{eqnarray}}
\newcommand{\eea}{\end{eqnarray}}
\renewcommand{\be}{\begin{equation}}
\renewcommand{\ee}{\end{equation}}
\newcommand{\rund}[1]{\left(#1\right)}
\newcommand{\vc}[1]{\mbox{\boldmath $#1$}}
\newcommand{\dc}{\partial}
\newcommand{\eck}[1]{\left[ #1 \right]}
\newcommand{\msun}{\,h^{-1}\,M_{\odot}}

\def\elabel#1{\label{eq:#1}}

\begin{document}
\title{Gravitational lensing properties of isothermal universal halo profile}

\volnopage{Vol.0 (200x) No.0, 000--000} 
\setcounter{page}{1}     

\author{Xinzhong Er
  \inst{1}
}
\institute{$^1$National Astronomical Observatories, Chinese Academy of Sciences,
  Beijing 100012, China; {\it xer@nao.ac.cn}\\
}
\date{Received~~; accepted~~}

\abstract{ N-body simulations predict that dark matter halos with
  different mass scales are described by a universal model, the
  Navarro-Frenk-White (NFW) density profiles. As a consequence of
  baryonic cooling effects, the halos will become more concentrated,
  and similar to an isothermal sphere over large range in radii ($\sim
  300$ $h^{-1}$kpc). The singular isothermal sphere model however has
  to be truncated artificially at large radii since it extends to
  infinity. We model a massive galaxy halo as a combination of an
  isothermal sphere and an NFW density profile. We give an
  approximation for the mass concentration at different baryon
  fractions and present exact expressions for the weak lensing shear
  and flexion for such a halo. We compare the lensing properties with
  a Singular Isothermal Sphere and NFW profiles. We find that the
  combined profile can generate higher order lensing signals at small radii
  and is more efficient in generating strong lensing events. In order
  to distinguish such a halo profile from the SIS or NFW profiles, one
  needs to combine strong and weak lensing constraints on small and
  large radii.
}
\authorrunning{Er} 
\titlerunning{Lensing of INFW halo} 
\maketitle
\section{Introduction}   
\label{sect:intro}

The Cold Dark Matter with the cosmological constant model
($\Lambda$CDM) provides a successful description of many properties of
observations of the universe. N-body simulations of $\Lambda$CDM
models predict dark matter halos with a universal density profile
\citep[e.g.][]{nfw97}. The Navarro-Frenk-White (NFW) profile appears
to be a good approximation for dark halo profiles over a wide range of
masses. On the other hand, the NFW halo density profile can also be
generalized with an arbitrary power law central cusp, and outer
regions that fall off as $r^{-3}$ \citep{2000ApJ...529L..69J}. It also
has been found that the slope of the inner regions steepens for
smaller mass haloes. More importantly, baryonic cooling will
significantly steepen the density profiles, close to the isothermal
slopes observed \citep{2009ApJ...703L..51K}. The baryon effect is more
significant in the galaxy halo since it contains more baryons. A
composite model with an NFW dark matter halo and a de Vaucouleurs
stellar component is suggested for massive galaxies by
\citet{2007ApJ...667..176G}. The total density profile is close to
isothermal form over large range in radius ($\sim300$ $h^{-1}$kpc).
Therefore, we model the halo total mass profile as an Isothermal-NFW
(INFW) profile, which is the combination of an NFW dark halo plus a
stellar component at inner radii, i.e. $\rho\propto r^{-2}$ for small
radius.

Gravitational lensing provides a direct way to study the mass
distribution of large scale structures in the universe as well as
galaxy and cluster halos. It probes the mass distribution independent
of the nature of matter or its dynamical state
\citep[e.g.][]{2001PhR...340..291B,2010ARA&A..48...87T}. Lensing is
widely used for the cluster mass reconstruction
\citep[e.g.][]{2006ApJ...652..937B}, and galaxy halo measurement
\citep[e.g.][]{2009MNRAS.394..929C}. 
Weak lensing is the physical phenomenon causing the weak image distortion
of background galaxies. By comparing the image distortions
with non-lensed image shapes, one can infer the mass distribution of
the foreground lens. In weak lensing, most studies consider the shear
effect, which transfers a round source into an elliptical one. Higher
order effects, flexion are gradually coming within reach. Flexion can
be introduced as derivatives of either the surface mass density or the
shear. They respond to smaller-scale variations in the projected mass
distribution than the shear \citep{bacon2006}. The
convergence gradient, called the first flexion ${\cal F}$, introduces
a centroid shift in the lensed image and is a spin-1 symmetry
quantity, while the second flexion ${\cal G}$ is the gradient of shear
and is spin-3. Flexion provides a measure of small scale
variations of mass distribution as well as the halo ellipticity
\citep{2011A&A...528A..52E,2012MNRAS.421.1443E}.

The lensing properties of different halo profiles have been widely
studied, e.g. the NFW profile \citep{1996A&A...313..697B} and the
Einasto profile \citep{2012A&A...546A..32R}.
\citet{2001ApJ...555..504W} and \citet{2001ApJ...549L..25K} have studied a
generalized NFW type profile for lensing. Therefore, it is
interesting to use the INFW profile as a galaxy halo, and it is natural to
extend its applications to the gravitational lensing characteristics of
dark matter halos. For first time, we apply analytical and numerical
methods to the gravitational lensing study of INFW halo profiles. In
Sect.\ref{sect:infw}, we present the basic halo properties of the INFW
profile. In Sect.\ref{sect:lens}, the analytical formula of an INFW lens
halo is given. We compare the INFW profile with other models in
Sect.\ref{sect:comp} and give a summary at the end.
The cosmology that we adopt in this paper is a $\Lambda$CDM model with
parameters based on the results of the Wilkinson Microwave Anisotropy
Probe seven year data \citep{2011ApJS..192...18K}:
$\Omega_{\Lambda}=0.734$, $\Omega_{\rm m}=0.266$, Hubble constant
$H_0 = 100 h$ km\,s$^{-1}$\,Mpc$^{-1}$ and $h=0.71$.

\section{INFW halo properties}
\label{sect:infw}

In analogy to NFW model, the density profile of INFW is given by
\be
\rho(r) = {\rho_{\rm c} \Delta'_c r_s^3\over r^2 (r_s + r)},
\ee
where $\rho_{\rm c}=3H(z)^2/(8\pi G)$ is the critical density of the
universe, $H(z)$ is the Hubble parameter, and $G$ is Newton's
constant. The dimensionless characteristic density is given by
\be
\Delta'_c= \frac{200 c_I^3}{3 {\rm ln}(1+c_I)}
\ee
\citep[see, e.g., ][]{2001ApJ...555..504W}. We will use the same
definition for the concentration $c_I=r_{200}/r_s$, where $r_s$ is the scale
radius. The virial radius $r_{200}$ is defined as the radius inside
which the mass density of the halo is equal to $200\rho_c$
\citep{nfw97}. The mass of a halo contained within a radius of
$r_{200}$ is thus
\be
M_{200} = {800 \pi \over 3} \rho_{\rm c} r_{200}^3.
\ee

\begin{figure}
\centerline{\scalebox{1.0}
{\includegraphics[width=6.5cm,height=6.0cm]{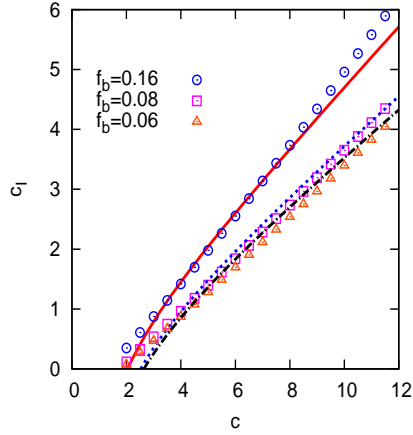}}}
\caption{The approximate relationship between the concentration $c$ for
  the NFW profile and $c_I$ for the INFW profile. The points are
  the numerical results from solving Eq.(\ref{eq:c2c}) for different baryon
  fractions: $f_b=0.16$ (circles), $f_b=0.08$ (squares) and $f_b=0.06$
  (triangles). The lines are the approximate relationship using
  Eq.(\ref{eq:appci}).}
\label{fig:c2c}
\end{figure}

\begin{figure}
  \centerline{\scalebox{1.0}
    {\includegraphics[width=6.5cm,height=6.0cm]{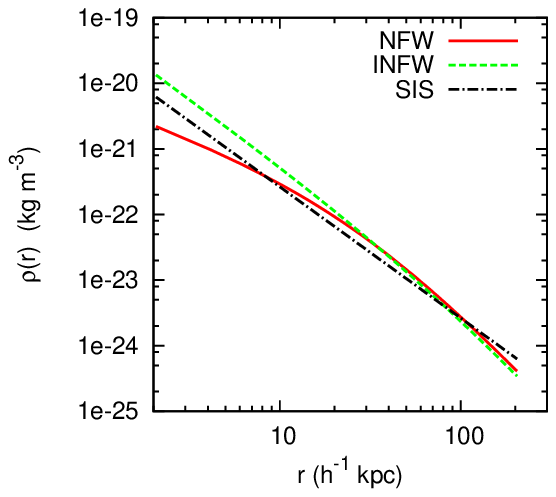}
      \includegraphics[width=6.5cm,height=6.0cm]{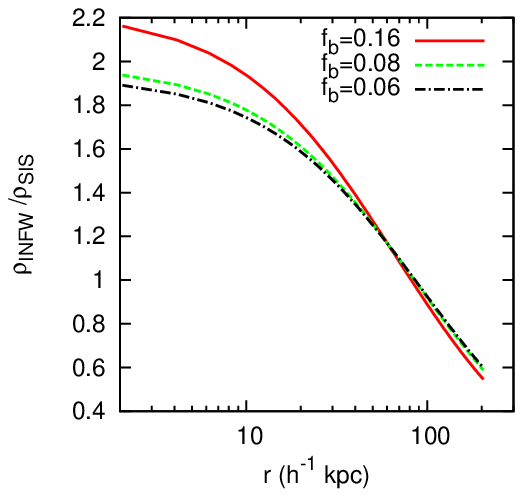}}}
  \caption{Left panel: halo mass density $\rho(r)$ for three different
    profiles: NFW (solid line), INFW (dashed line), SIS (dot-dashed
    line). The same mass ($M_{200}=10^{12} \msun$) is used for
    different profiles (also for right panel). The concentration is
    $c=6.95$ ($c_I=3.11$, $f_b=0.16$) for NFW (INFW) halo. Right
    panel: $\rho(r)_{\rm INFW}/\rho(r)_{\rm SIS}$ with different
    $f_b$: 0.16 (solid line), 0.08 (dashed line), 0.06 (dot-dashed
    line).}
  \label{fig:rp}
\end{figure}

There is no specific study or simulations for the relationship between mass
and concentration for generalized-NFW profiles.
We assume that initially dark matter and baryons follow the same NFW
profile. Due to the cooling effect, baryons collapse toward the center of the
dark matter halo, and steepen the inner density profile. We assume
the collapsed baryons make a fraction $f_b$ of the total mass. The
outer density of the INFW profile will become lower by a factor of
($1-f_b$): $\rho_{\rm INFW}(r_{200}) = (1-f_b) \rho_{\rm NFW}(r_{200})$. We
take $f_b$ as the universal baryon fraction, although a lower number
does not change the scaling significantly. A relation between $c$ and
$c_I$ can be obtained from
\be {c_I\over {\rm ln}(1+c_I)\,(1+c_I)} =
(1-f_b){c^2 \over \eck{{\rm ln}(1+c)- {c\over 1+c}} (1+c)^2}.
\elabel{c2c}
\ee
This relation can be solved numerically. It can be also approximated
by
\be
c_I = {c \over 3-6 f_b} - {3-6 f_b \over c}.
\elabel{appci}
\ee
In Fig.~\ref{fig:c2c}, one can see that our approximation mainly
agrees with the numerical results. A smaller baryon fraction will lead
to a lower concentration of the INFW halo. When the concentration
$c_I$ becomes to $0$, $r_s\to \infty$, and the INFW profile reduces to
an SIS. Thus in general the INFW profile is more concentrated than SIS
profile at small radial. We will see in next section that the INFW
profile can produce higher lensing signals and is more efficient in
generating strong lensing than other profiles. The small variation of
baryon fraction does not strongly affect the matter density profile
(right panel of Fig.\,\ref{fig:rp}). With higher baryon fraction, the 
density at inner radius is larger. In the rest of the paper, we will
use $f_b=0.16$ and Eq.(\ref{eq:appci})to estimate the concentration of
the INFW halo.

In left panel of Fig.\,\ref{fig:rp} we show $\rho(r)$ for three
different halo profiles using same halo mass $M_{200}$. One can see
that the INFW profile has the same slope as SIS at small radii ($<30$
$h^{-1}$ kpc) and approaches to NFW at large radii.


\section{Lensing properties of INFW halo}
\label{sect:lens}
\subsection{Basic lensing formula}
The fundamentals of gravitational lensing can be found in
\citet{2001PhR...340..291B}. For its elegance and brevity, we shall
use the complex notation. The thin-lens approximation is adopted,
implying that the lensing mass distribution can be projected onto the
lens plane perpendicular to the line-of-sight. We introduce angular
coordinates $\vc\theta$ on the lensing plane with respect to the
line-of-sight. The lensing convergence, that is the dimensionless
projected surface-mass density, can be written as
\be
\kappa(\vc\theta) = \Sigma(\vc\theta)/\Sigma_{\rm cr},\;\;\; {\rm where} \;\;\;
\Sigma_{\rm cr} = \frac{c^2}{4\pi G} \frac{D_{\rm s}}{D_{\rm d} D_{\rm ds}}\;
\ee
is the critical surface mass density depending on the angular-diameter
distances $D_{\rm s}$, $D_{\rm d}$ and $D_{\rm ds}$ from the observer
to the source, the observer to the lens, and the lens to the source,
respectively. $\Sigma(\vc\theta)$ is the projected surface-mass
density of the lens. All lensing quantities can be derived from the
effective lensing potential $\psi$,
\be
\psi(\vc\theta) = \frac{1}{\pi}\int_{{\cal R}^2} \d^2\theta'\kappa(\vc\theta')\;
    {\rm ln}|\vc\theta-\vc\theta'|\;.
\ee

The lens equation is given by
\be
\beta= \theta - \alpha(\theta),
\ee
where $\beta$ is the source position and $\alpha$ is the deflection angle
\be
\alpha = \nabla_c \psi, 
\ee
where the complex differential operators is defined as
\be
\nabla_c:= {\dc \over \dc \theta_1 } + \ii{\dc \over \dc \theta_2};\quad\quad
\nabla_c^*:={\dc \over \dc \theta_1 } - \ii{\dc \over \dc \theta_2}.
\ee
To the lowest order, image distortions caused by gravitational lensing
are described by the complex shear and convergence (which equals to
the dimensionless surface mass density)
\be
\gamma = \frac{1}{2}\left(\partial_1^2\psi-\partial_2^2\psi\right)
+ {\rm i}\partial_1\partial_2\psi\; = {1\over2 } \nabla_c^2 \psi; \quad
\kappa = {1\over2}(\dc^2_1\psi+\dc^2_2\psi) = {1\over2} \nabla_c\nabla_c^*\psi,
\label{shear}
\ee
where the subscripts $i$ denote partial derivatives with respect to $\theta_i$.
The magnification for a point source is thus given by
\be
\mu = \dfrac{1}{(1-\kappa)^2 \,-\,|\gamma|^2}.
\elabel{anamu}
\ee
The shear transforms a hypothetical round source into an elliptical image.
The ${\cal F}$ and ${\cal G}$ flexions can be introduced as the complex
derivatives
\be
\mathcal{F} = \nabla_c \kappa; \quad\quad
\mathcal{G} = \nabla_c\gamma.
\label{eq:1c}
\ee
The flexions are thus combinations of third-order derivatives of the
effective lensing potential $\psi$. We shall denote their real and
imaginary parts by $(\mathcal{F, G})_1$ and $(\mathcal{F, G})_2$,
respectively. In terms of the lensing potential, we have
\begin{equation}
  \mathcal{F} \equiv {\cal F}_1 + \ii {\cal F}_2
= \frac{1}{2}\left(\partial_1^3\psi + \partial_1\partial_2^2\psi \right)+\frac{\ii}{2}\left(\partial_1^2\partial_2\psi + \partial_2^3\psi\right)
\label{eq:1d}
\end{equation}
and
\begin{equation}
  \mathcal{G} \equiv {\cal G}_1 + \ii {\cal G}_2
= \frac{1}{2}\left(\partial_1^3\psi - 3\partial_1\partial_2^2\psi \right)+\frac{\ii}{2}\left(3\partial_1^2\partial_2\psi-\partial_2^3\psi\right)\;.
\label{eq:1e}
\end{equation}

\subsection{Lensing of INFW halo}
We derive the analytical expression for the lensing properties of INFW halo.
The surface mass density of a spherically symmetric lens is obtained by
integrating along the line of sight of the three-dimensional density profile,
\be
\Sigma(\xi)= \int_{-\infty}^{\infty} \rho(\sqrt{\xi^2+z^2}) \d z,
\ee
where $\xi$ is the distance from the center of the lens in the
projected lens plane $\xi=\theta D_{\rm s}$. It implies the following form
for the dimensionless surface mass density
\be
\kappa(x) = 2 \kappa_s \rund{{\pi \over 2x} - f(x)},
\ee
where $x=\theta/\theta_s\,(\theta_s=r_s/D_{\rm d})$, and $f(x)$ is given by
\be
f(x)=
\begin{cases}
  \dfrac{{\rm arcsech} x }{\sqrt{1 - x^2}} \quad\;\; (x<1); \\ \\
  1 \quad\quad\quad\quad\quad (x=1); \\ \\
  \dfrac{{\rm arcsec} x }{\sqrt{x^2 - 1}} \quad\;\; (x>1).\\
\end{cases}
\label{eq:fx}
\ee
In the spherical case, the deflection angle is given by
\be
\alpha(\theta) = {2\over \theta} \int^{\theta}_0 \theta \d \theta \kappa(\theta)
\;=\; {4\kappa_s \theta_s \over x}\rund{{\pi x\over 2} +
(1-x^2)f(x) + {\rm ln}{x \over 2}}.
\ee
The analytical form of the shear can be
calculated from $\gamma(\theta)=\eck{\bar\kappa(\theta)-\kappa(\theta)}
{\rm exp}[2\ii\phi]$, where $\phi$ is the polar angle.
$\bar\kappa(\theta)$ is the mean surface mass density within a
circle of radius of $\theta$ from the lens center \citep[see
  e.g.][]{2001PhR...340..291B}. The expression for shear due to the INFW is
\be
\gamma(x) =
2\kappa_s \eck{{\pi\over 2x} + \dfrac{2 {\rm ln}(x/2)}{x^2} +
  {2-x^2\over x^2}f(x)} {\rm e}^{2\ii \phi},
\ee
where $f(x)$ is defined in Eq. (\ref{eq:fx}).
The analytical form of two flexions can be also given
\bea
{\cal F}(x) &=& {2\kappa_s \over \theta_s} \eck{{xf(x) \over x^2-1}
- {\pi \over 2 x^2} -{1 \over x (x^2-1)}} {\rm e}^{\ii \phi},\\
{\cal G}(x) &=& {2\kappa_s\over \theta_s}
\rund{-{3\pi\over 2x^2} -{8{\rm ln}(x/2)\over x^3} + {1\over x(x^2-1)} -f(x)
\eck{{8\over x^3 } - {3\over x} +{1\over x(x^2-1)}}} {\rm e}^{3\ii \phi}.
\eea
The elliptical INFW lensing properties can be calculated numerically
\citep{2001astro.ph..2341K}.

Furthermore, as pointed out by
\citet{1995A&A...294..411S,schneider&er08}, due to the mass-sheet
degeneracy, the directly measurable properties are the reduced shear and
reduced flexion
\be
g = {\gamma \over 1-\kappa};\;\;\;
G_1 = {{\cal F} + g {\cal F}^* \over 1-\kappa };\;\;\;
G_3 = {{\cal G} + g {\cal G} \over 1-\kappa }.
\ee
The weak lensing properties of the INFW profile also show approximated
behavior as a combination of two power-law profiles. At small radii, the
asymptotic behavior can be approximated by a SIS, i.e.  $\kappa,\,\gamma
\propto \theta^{-1}$, and ${\cal F},\,{\cal G} \propto
\theta^{-2}$. At large radii, it behaves like the power-law
$\rho\propto r^{-3}$. Thus the lensing signal rapidly fade out,
$\kappa,\,\gamma \propto \theta^{-2}$, and ${\cal F},\,{\cal G}
\propto \theta^{-3}$.

\section{Profiles comparisons}
\label{sect:comp}
We compare the weak lensing properties for INFW, NFW, and SIS
profiles. We use an approximation relation to calculate the mass
concentration of the NFW profile \citep{2007MNRAS.381.1450N}
\be
c= 5.26\rund{M_{200} \over 10^{14} \msun}^{-0.1},
\ee
and use Eq.\ref{eq:appci} to obtain $c_I$.
The velocity dispersion of the SIS profile $\sigma_v$ is calculated through
$\sigma_v^3 = {5 \over \sqrt{2}} G\, H(z)\, M_{200}$
\citep{1998MNRAS.295..319M}.
The lensing properties of the SIS or NFW profiles can be found in e.g.
\citet{2000ApJ...534...34W, 2006glsw.book.....S}.
We use lens halo mass $M_{200}=10^{12}\msun$, which is a galactic sized
halo. The lens is placed at redshift $z_{\rm d}=0.2$ and the sources are at
redshift $z_{\rm s}=1.0$, which are accessible median redshifts for galaxy
survey, e.g. SDSS or LSST. The concentration parameter for the NFW
(INFW) profile in our test is $c=6.95$ ($c_I=3.11$). The Einstein
radius of the SIS profile is $\theta_{\rm E}\approx 0.3$ arcsec.

\begin{figure}
  \centering{
    \includegraphics[width=60mm,height=56mm]{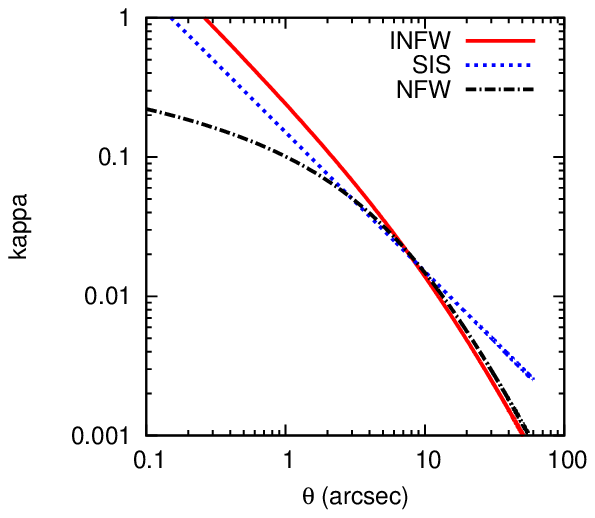}
    \includegraphics[width=60mm,height=56mm]{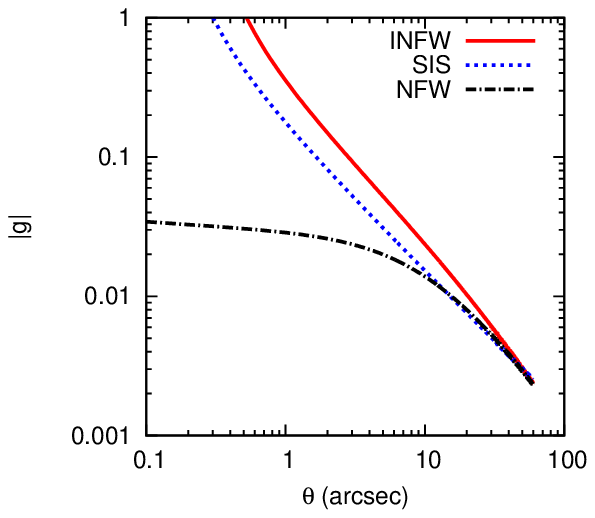}}\\
  \centering{
    \includegraphics[width=60mm,height=56mm]{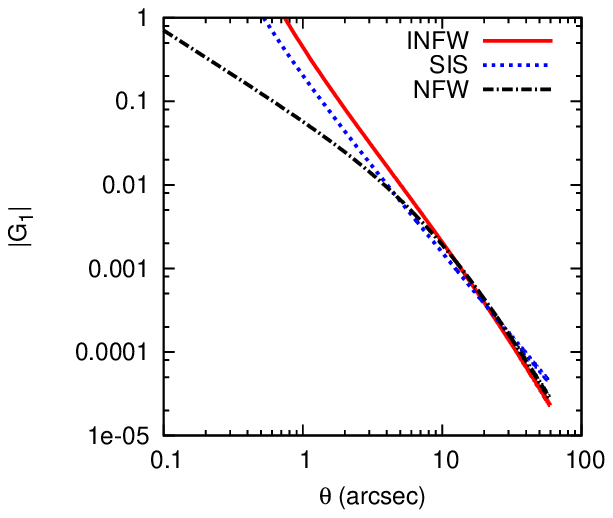}
    \includegraphics[width=60mm,height=56mm]{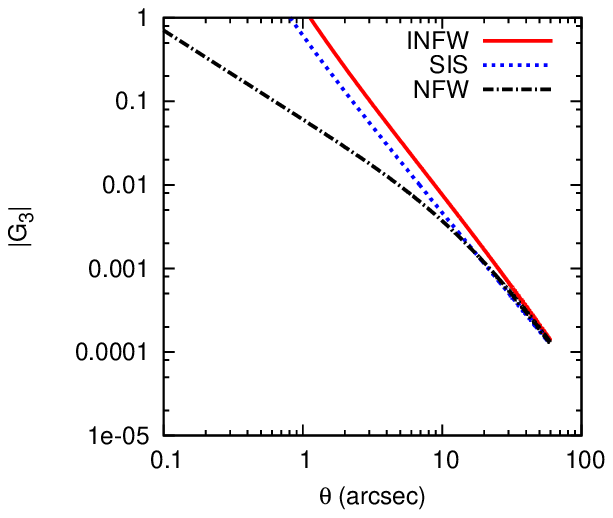}}
  \caption{Convergence, reduced shear and reduced flexions of three different
    profiles: INFW (solid line), SIS (dashed line), NFW (dot-dashed
    line). The mass of the lens halo is $M_{200}=10^{12}\msun$. The
    lens and source redshifts are assumed to be $z_d=0.2$ and $z_s=1.0$
    respectively.}
  \label{fig:4comp}
\end{figure}

Fig.~\ref{fig:4comp} shows the predicted convergence, reduced shear, 
first and second reduced flexions as a function of the angular
separation from the lens center. The mass profile of the mock galaxy
halo is assumed to be SIS (dotted line), NFW (dot-dashed line) and
INFW (solid line) model.
One can see that the overall behaviors of the three profiles are
comparable. The asymptotic lensing behavior of the INFW profile are
proportional to that of the SIS profile at small radii and approach
NFW profile at large radii. The signal magnitudes of all lensing
properties for the INFW halo are stronger than the other two at small
radii but drop faster and eventually below that of the other two
profiles. The differences between the magnitudes of the lensing signal
are stronger at small radii than that at large radii. In particular,
the shear and second flexion show a great dissimilarity. At large
radii, the difference between three profiles is not significant.

\begin{figure}
\centerline{\scalebox{1.0}
{\includegraphics[width=6.5cm,height=5.8cm]{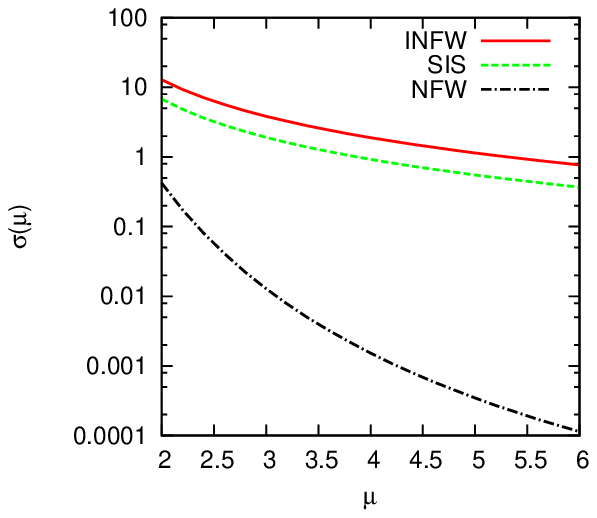}
  \includegraphics[width=6.5cm,height=5.8cm]{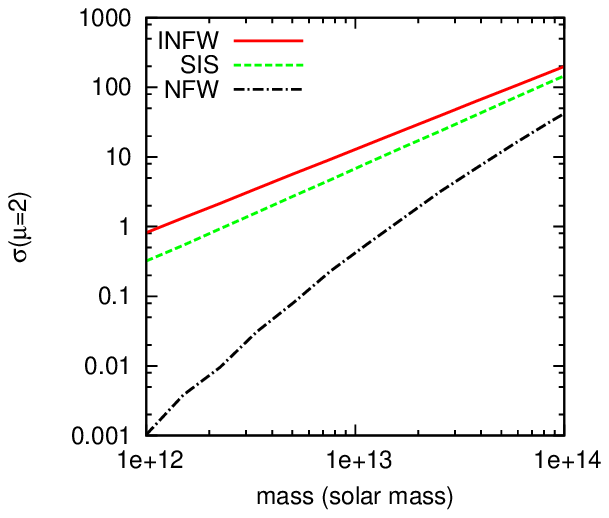}}}
\caption{The cross section of lensing magnification $\sigma(\mu)$ (in
  unit of arcsec$^2$) for three halo profiles: INFW (solid line), SIS
  (dashed line), NFW (dot-dashed line). The left panel shows the cross
  section as a function of lensing magnification for a lens halo with
  mass $M_{200}=10^{13}\msun$. The right panel shows the cross section
  of given magnification ($\mu=2$) for different lens halo mass. Same
  redshift condition of Fig.\ref{fig:4comp} ($z_{\rm d}=0.2$, $z_{\rm s}=1.0$) is
  used. }
\label{fig:sigma}
\end{figure}

In particular, the weak lensing properties of NFW halos is
significantly different from the other two profiles at small radius
($<10$ arcsec). The lensing signal of NFW is shallower than the
others. On the other hand, the signal magnitude of INFW halo is about
$2$ times higher than that of SIS halo. The lensing signals of INFW
drop faster with increasing radius, as one can see clearly from the
first flexion (the convergence $\kappa$ is not an observable
quantity). Moreover, the angular separation at which the INFW halo
first flexion is exceeded by other profiles is around $10$ arcsec, and
it is larger for shear and second flexion.
In principle one can study the weak lensing signal, i.e. the shear and
flexion to constrain the halo density profile. However, the weak
lensing signal at large radii is small and hard to detect. On the
other hand, it is also difficult to measure weak lensing signal when
the background image is close to the lens galaxy. One can perform
stacking method for galaxy-galaxy lensing studies. A large volume
survey is necessary.

Moreover, the significant difference lensing properties at small
radius will cause different strong lensing signal. In order to simply
see the strong lensing properties, we compare the magnification cross
sections for the three profiles. The cross section for a given
magnification threshold is defined as
\be \sigma(\mu_{\rm min}) = \int_{|\mu|>\mu_{\rm min}} \d^2\beta =
\int_{|\mu|>\mu_{\rm min}}{1\over |\mu|} \d^2\theta.  
\ee
In Fig.~\ref{fig:sigma}, the magnification cross section is shown in
the left (right) panel for halo with mass $10^{13}\msun$ (different
halo mass). We can see from both panels that the INFW profile can
generate larger cross sections than the other two profiles, due to the
high mass concentration of the INFW profile (top left panel in
Fig.~\ref{fig:4comp}). The cross section of NFW halo increases
faster with mass than other profiles, but decrease faster with
$\mu$. The curves of INFW and SIS profiles again have similar shapes, but
the cross section of INFW halo is about two times larger than that of
SIS halo for halo mass of $10^{13}\msun$.
In additional tests, we also study the cross section of strong lensing
multiple images. The probability generated by INFW halo can be several
times higher than NFW halo, and will be easy to distinguish from each
other. On the other hand, the INFW model generates about $3$ times
higher multiple image cross section than SIS model with halo mass
$\sim10^{12}\msun$ and approaches to that of SIS model for massive
halo ($>10^{15}\msun$). The concentration $c_I$ becomes small for
massive halo, thus the INFW nearly reduces to SIS profile. The
multiple image separation generated by INFW lens can reach $4$ arcsec
for a halo mass of $10^{13}\msun$, which is about $40$ percent larger
than that generated by SIS lens. Therefore, the galaxy-galaxy strong
lensing statistics can be a potential tool to distinguish INFW and SIS
profile.

\section{Summary}
We have studied the lensing properties of the INFW mass profile. The
INFW profile is motivated by the combination of Cold Dark Matter
simulations and a stellar component in the inner region of the dark
matter halo, together with some evidence from observations
\citep{2007ApJ...667..176G}. The inner profile of INFW is isothermal,
i.e. $\rho\propto r^{-2}$ and the outer profile is NFW-like $\rho\propto
r^{-3}$. An approximate mass concentration due to different baryon
fractions is given for the INFW profile, as a direct consequence of
baryon collapse toward the center of halo.

The analytical expressions for deflection angle, convergence, shear and
flexions of an INFW halo lens are given. We have compared the lensing
properties of INFW profile with NFW and SIS halo profiles. We find
that the INFW profile is more efficient than the others in generating
lensing magnification, and the weak lensing signals of INFW halo is
stronger at small radii than that of other profiles for the same halo
mass. Strong lensing statistics can be used to constrain the lens
profile, e.g. the image separation.  However, the image separation
statistics is only sensitive to the inner profile of the lens
halo. There is a degeneracy between the massive SIS lens and high
concentrated INFW lens. Therefore, using weak lensing to study the
large radial profile is essentially necessary.

\begin{acknowledgements}
I thank Shude Mao, Ismael Tereno and Richard Long for discussions and
comments on the manuscript. I also thank referee for comments and
corrections on the manuscript. 
\end{acknowledgements}


\bibliographystyle{raa}
\bibliography{../../../bib/refbooks,../../../bib/lens,../../../bib/bhlens,../../../bib/refcos,../../../bib/flexion,../../../bib/galaxy,../../../bib/stronglens}

\end{document}